# Machine learning for classifying and interpreting coherent X-ray speckle patterns

Mingren Shen[1], Dina Sheyfer[2], Troy David Loeffler[3], Subramanian K.R.S. Sankaranarayanan[3,5], G. Brian Stephenson[4], Maria K. Y. Chan[3], Dane Morgan[1]

[1] Department of Materials Science and Engineering, University of Wisconsin-Madison, Madison, Wisconsin, 53706, USA

[2] X-ray Science Division, Argonne National Laboratory, Lemont, IL 60439, USA

[3] Center for Nanoscale Materials, Argonne National Laboratory, Lemont, Illinois 60439, USA

[4] Materials Science Division, Argonne National Laboratory, Lemont, IL 60439, USA

[5.] Department of Mechanical and Industrial Engineering, University of Illinois, Chicago IL – 60607, USA

**Abstract**

Speckle patterns produced by coherent X-ray have a close relationship with the internal structure of materials but quantitative inversion of the relationship to determine structure from speckle patterns is challenging. Here, we investigate the link between coherent X-ray speckle patterns and sample structures using a model 2D disk system and explore the ability of machine learning to learn aspects of the relationship. Specifically, we train a deep neural network to classify the coherent X-ray speckle patterns according to the disk number density in the corresponding structure. It is demonstrated that the classification system is accurate for both non-disperse and disperse size distributions.

# 1. Introduction

Current and developing X-ray sources such as the advanced synchrotron sources, X-ray free electron lasers, and high harmonic generation sources[1,2] enable utilization of coherent X-ray light to investigate behavior in the time domain and structure at interatomic length scales. For example, X-ray photon correlation spectroscopy has been used by many researchers in materials science, including investigations of micro, nano, and atomic scale structures[3,4] and mechanisms[5], studying dynamics and correlation behaviors[6–8], revealing rich phenomena in complex material systems, e.g. multicomponent fluids[9] and metallic glasses[10], and in other areas. Coherent X-ray imaging methods have been used in many material applications to visualize the chemical composition at nanoscale resolution[11,12] and to study 3D lattice dynamics in nanocrystals[13], and in biology to image the 3D mass density distribution of a whole cell[14], and to reconstruct the 3D structure of the giant mimivirus particle[15].

Coherent X-rays incident on a disordered sample generate X-ray speckle patterns, which are often difficult to interpret, especially to reconstruct molecular structure from the speckle patterns. Such reconstructions to date have often relied on complex, subjective algorithms or required multiple experiments, e.g., using phase retrieval algorithms to iterate between real and reciprocal space[16,17] or alternating projections[18], angles or sample positions[19]. The reconstruction problem occurs because only amplitude information is recorded in the detector and additional computation is required to recover phase information. Deep learning, if used properly, has been shown to be able to learn complex mapping between input space and output space automatically from data[20] and has been widely used for physics[21–24], chemistry[25], material science[26–29], health industry[30–32], etc. It is therefore of interest to explore whether recent developments in image analysis using deep learning might aid in the extraction of structural information from X-ray speckle patterns.

Machine learning or deep learning methods can help X-ray speckle pattern problems in three aspects, namely data collection, data transformation, and data analysis. For data collection, the main goal is to acquire high resolution and lower noise images. Konstantinova et al. use a convolutional neural-network-based encode-decoder framework to reduce noises in X-ray images[33] and Cherukara et al. use neural networks to boost low resolution scanning coherent diffraction images to high fidelity ones[34]. Other ideas like developing a novel machine-learning-based data acquisition process to interpret common nanoscale lattice structural distortions[35] and using deep reinforcement learning to help lower dose used for ptychography experiments[36] also show great promise. For data transformation, the aim is preprocessing raw data to get prepared data for analysis. For example, neural-network-based data augmentation methods are used to increase data volume[37,38], and physics information such as geometry[39] and spherical harmonics for representing crystals[40] are incorporated into machine learning methods to get better performance. A machine-learning-based automatic X-ray scattering image annotation system has also been developed by Guan et al. to help increase the speed of data ingestion[41]. For data analysis, more powerful machine learning models[42–44], better features[45], and new templates[46] have all been attempted for various tasks showing promising advantages compared to traditional methods[34–36,38–42].

For this study, the major challenge impeding the application of supervised machine learning or deep learning methods for X-ray speckle pattern problems[47] is a lack of labeled data for training, since we generally have little or no knowledge of ground truth values for the real physical systems being characterized. Two general approaches to overcoming this challenge are (i) to use synthetic data, e.g., generated by forward simulation of relevant molecular structures, and (ii) to make extremely efficient use of data that is available through approaches like transfer learning, few shot learning, physics guided models, and data augmentation. The two approaches can certainly be used together. We take approach (i) in the present work, demonstrating that we can effectively use machine learning to extract structural information from simulated speckle patterns derived from simplified forward simulation on a very idealized model system. However, this is a very preliminary step, and a combination of more physically realistic synthetic data, experimental data, and highly data efficient machine learning approaches will likely be needed to realize practical models. To better focus on examining machine learning model's ability to interpret speckle patterns, we use a simple 2D disk system to generate speckle patterns, however, more realistic synthetic data of speckle patterns could be obtained from better material simulation methods such as Molecular Dynamics (MD) or Kinetic Monte Carlo (KMC)[48–50].

In this study, we explore the ability of machine learning to learn aspects of the relationship between coherent X-ray speckle patterns and sample structures using a model 2D disk system. Specifically, we train a deep neural network to classify the coherent X-ray speckle patterns according to the disk number density in the corresponding structure. Disk number density is one of the simplest and most direct properties to determine and is analogous to number densities often of interest in colloidal systems. We therefore consider it a good property for an initial exploration of the capabilities of deep learning for this problem. We focus on a representative 2D system to allow a more rapid study at this initial stage, although extension to 3D systems is straightforward and clearly a valuable focus for future work. The classifier was trained on simulated coherent X-ray speckle patterns without any assumptions or simplifications and the results suggest that it is possible to use machine learning tools to correlate coherent X-ray speckle patterns to structural or particle distribution information with machine learning. The method directly builds a mapping between the $k$ space coherent X-ray speckle patterns and real space sample information, e.g., disk number density. The present model is trained on synthetic data for an idealized system and therefore not directly applicable to speckle patterns from physical samples. However, its success suggests that, if accurate training data can be obtained, a deep learning algorithm could be developed to aid in structural analysis from speckle patterns in real systems. In the discussion, we compare the results to a conventional X-ray scattering analysis.

## 2. Material and methods

**Overview**

We used forward-simulation methods to generate coherent X-ray speckle patterns of a model 2D disk system and then trained a Convolutional Neural Network (CNN) model called ResNet-50 to classify the X-ray speckle patterns into different categories corresponding to their disk numbers. The workflow pipeline of this approach is shown in Figure 1. The generated speckle patterns were used as input and the disk number as the target output class for neural network. These were then used to train the ResNet-50 network to classify a speckle pattern by its number density.

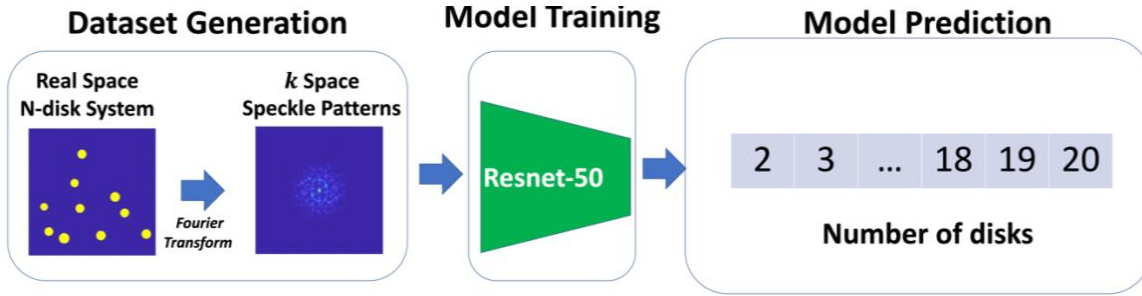

*Figure 1. Flow chart of machine learning system which includes dataset generation, model training and model prediction and interpretation.*

**Model system generation**

The 2D system used in our study was $2000 \times 2000$ units in real space and the total area of disks is kept at a constant (equal to $2 \times \pi \times 90 \times 90 = 16200\pi$ where 90 is the radius of the two-disk system). For an *n*-disk system, the radius of each disk was set to preserve the total area of disks, which requires $r = \sqrt{16200/n}$. Then the *n* disks were randomly placed into the 2D box without overlap of disks. For a select set of studies focused on polydispersity, the single radius system of *n* disks was modified by sampling the radii from a Gaussian distribution of a given mean and standard deviation with the constraint that the total area of disks is kept the same as the original single radius *n*-disk system ($16200\pi$). The standard deviation was chosen to be 1/3 of the mean to represent a significant but not overwhelming spread in particle sizes compared to the mean, as might be seen in a real nano-cluster system.

**Speckle pattern dataset generation**

To generate X-ray speckle pattern data, we use a Fourier Transform to covert the real space images of 2D *n*-disk system to the reciprocal *k* space speckle patterns. The width of each side of a pixel in *k* space is $w = \pi/L$ where *L* is the size of the real space images of disks. Each pixel in the X-ray speckle patterns is centered around the $k$ space coordinate used to calculate the Fourier transform for that pixel. The $k$ space coordinates are of the form ($n_x \times w$, $n_y \times w$), where $n_x$ and $n_y$ are integers ranging from 0 to $n_{max}$, and $n_{max}$ can be 256 or 512, yielding speckle patterns of size 257 × 257 and 513 × 513, respectively. The $k$ space speckle patterns range along each axis is from 0 to $k_{max}$, where $k_{max} = \frac{n_{max} \times \pi}{L}$. In Figure 2, we show real space and corresponding $k$ space speckle patterns for selected real space images of $n$-disk systems ($n$ = 2, 3, 6, 10, 15, 20). For clarity, all $k$ space speckle patterns are arbitrarily scaled to highlight the speckle patterns details. In total, we consider 19 cases given by $n$ = 2 - 20.

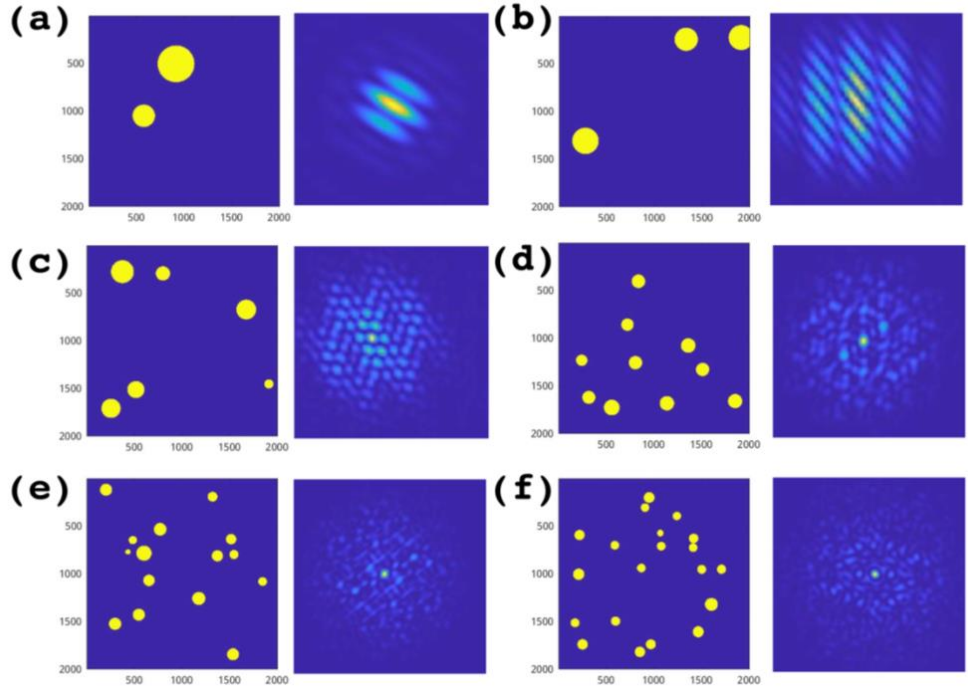

*Figure 2. Real space images of n-disk system and corresponding X-ray speckle patterns, arbitrarily scaled to highlight the speckle patterns for six typical 2D disk system with (a) 2-disks, (b) 3-disks, (c) 6-disks, (d) 10-disks, (e) 15-disks, and (f) 20-disks.*

**Data processing**

After obtaining the X-ray speckle patterns, some preprocessing was performed. The 11 × 11 pixels center region of the X-ray speckle patterns was blocked to mimic what is commonly done in experiments to avoid the excessive brightness of the center pixels[19]. Standardization was applied to each speckle pattern to make sure the pixel intensity of all speckle patterns has the zero mean and unit variance. This is an important step to remove any correlation between the overall intensity and the number density of disks to ensure the machine learning system is forced to focus on the structure of X-ray patterns rather than the intensity distribution. This standardization mimics what would be needed in real data where the contrast of the objects is typically unknown.

**Machine learning model and data: properties, fitting, and analysis**

We applied the widely used CNN model called ResNet-50 to classify the speckle patterns based on the number of disks. ResNet[51,52] is a family of deep learning models that uses identity mappings to overcome the performance degradation problem of stacking more layers[53]. It is the first deep learning model that achieved lower than human level error rate in the ImageNet Large Scale Visual Recognition Challenge (ILSVRC) 2015 competition[54]. For this work, we created 1000 different random configurations of disk position each for 19 different disk numbers. In total, we have 19000 X-ray speckle patterns and among them, 15200 speckle patterns were used for training, 1900 speckle patterns were used for validation, and the remaining 1900 speckle patterns were used for testing. For polydisperse system, the 1900 speckle pattern test set was randomly selected for convenience, so different disk numbers may have different number of testing speckle patterns. And for non-polydisperse system, the 1900 speckle pattern test set contains exactly 100 speckle patterns for each disk number 2 - 20 to keep testing unbiased for disk numbers. This condition was enforced by randomly selecting 100 speckle patterns from the 1000 total speckle patterns of each disk number to build the test set. The training and validation set were randomly split from the remaining 17100 speckle patterns after the test set was extracted. ResNet-50 model was written in Keras[55] with TensorFlow[56] as the backend engine and the training used the Adam optimizer and categorical cross entropy loss with default setting of Keras. A typical training used 200 epochs with batch size of 40.

To estimate the performance of our models, we will use precision, recall, F1 scores and confusion martix. Here precision means the ratio between correct classified speckle patterns over all predicted speckle patterns, recall means the ratio between correct classified speckle patterns over all testing speckle patterns, and F1 scores is the harmonic mean of precision and recall. A confusion matrix is a widely used tool for a comprehensive evaluation of classifier performance. Essentially, it juxtaposes the predicted classes against the true classes, culminating in an $N \times N$ matrix. Diagonal entries in this matrix signify correct classifications for each class, whereas off-diagonal entries indicate misclassifications. This matrix not only facilitates a visual representation of accurate and erroneous predictions across classes but can also be readily used for computing True Positives (TP), True Negatives (TN), False Positives (FP), and False Negatives (FN). These values are pivotal for deriving critical classification metrics such as precision, recall, and F1-scores, providing a holistic view of a classifier's efficacy for each class.

### 3. Results

We first present the non-polydisperse system results where all disks were the same size in each real space images of $n$-disk system and then we show the polydisperse system results where disk sizes follow a Gaussian distribution (the construction of the polydisperse system was discussed in Section 2).

### 3.1 Non-Polydisperse System Results

The non-polydisperse classification results as confusion matrix on test data for 19 different disk numbers are shown in Figure 3. Figure 3 demonstrates that the classification algorithm works extremely well. Misclassification only happens once, in the case of 19 disks, which only shows a small error by misclassifying 19 as 18. This one error is the only misclassification in all 1900 test cases. The cross-class accuracy is nearly 99.9% which demonstrates the capacity of the present ResNet-50 based deep learning classification system to extract structural information from X-ray speckle patterns, at least when sufficient training data is available.

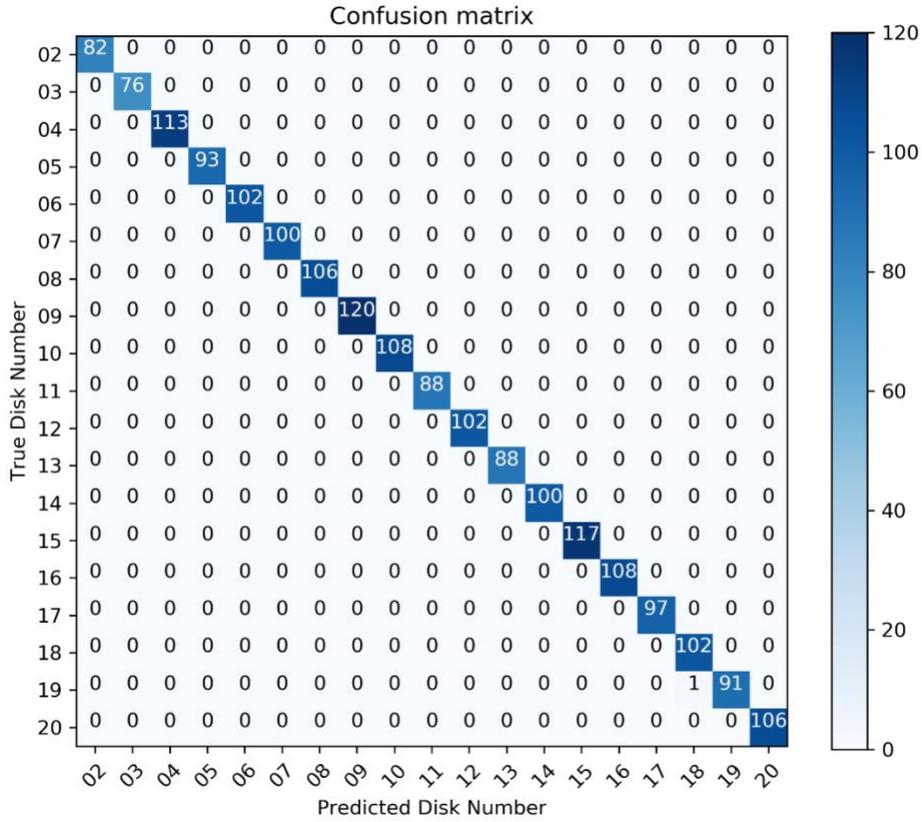

*Figure 3. Confusion Matrix of non-polydispersity system classification results using $257 \times 257$ as input size of coherent X-ray speckle patterns.*

In examples above, we focused on $257 \times 257$ pixel size as the coherent X-ray speckle pattern inputs, and it is important to know that whether different size of inputs will change the classification results. We increased the number of pixels from $257 \times 257$ to $513 \times 513$, which, following the *k* space scheme described in the Methods section, effectively doubled the range of *k* values sampled but kept the sampling density the same. This change added many higher *k* values, or equivalently, extended the sampling to smaller distances in real space (specifically, exploring the additional real space ranges of $2L/256$ down to $2L/512$. We applied the exact same neural network that was obtained from training on the $257 \times 257$ speckle patterns to classify the $513 \times 513$ speckle patterns. We show the confusion matrix of $513 \times 513$ size inputs in Figure 4, which shows few changes compared to Figure 3. The drop of performance corresponds to three misclassifications compared to just one from the smaller speckle patterns used previously, which is a very modest increase in error. This result indicates that the $257 \times 257$ speckle patterns reaches high enough $k$ values for our classification problem, even though the larger size inputs have more information.

In summary, the deep learning classifier works very well for identifying the number of disks from coherent X-ray speckle patterns in our model non-polydisperse systems.

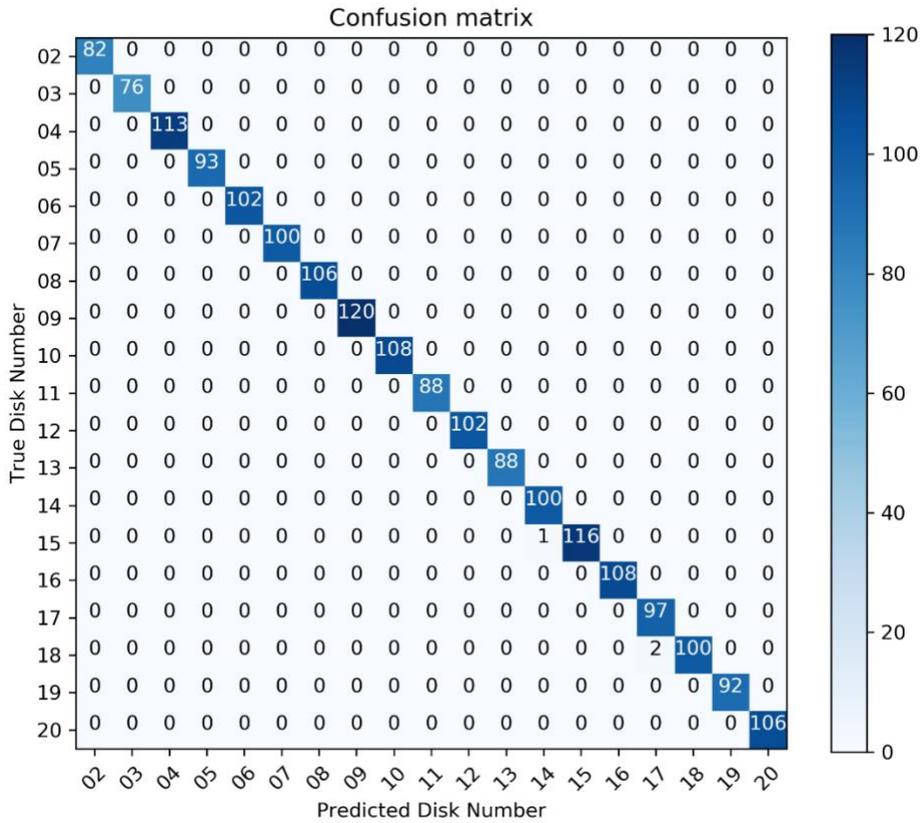

*Figure 4. Confusion Matrix of non-polydispersity system classification results using $513 \times 513$ as input size of coherent X-ray speckle patterns.*

**3.2 Polydisperse System Results**

In Section 3.1, we presented results of the deep learning classifier for coherent X-ray speckle patterns in non-polydisperse systems. However, real systems will always have some dispersion, so to further test the capacity of the classifier, we retrained the model and tested it on a polydisperse dataset (see Methods for how the polydisperse dataset was constructed). Figure 5 shows the confusion matrix of the polydisperse classifier. There is a significant performance drop compared to non-polydisperse results. The cross-class precision, recall and F1-score are all 0.89, lower than the non-polydisperse system due to the drop of performance in larger disk number systems from 14 to 20 as shown in Table 1 which shows class specific precision, recall, F1 scores. This decline in performance results from the added complexity of learning properties of a mixture of smaller and larger disks. One particular aspect that may be leading to issues is that as the size of the disks decreases, the distinction between these smaller disks becomes increasingly challenging due to their small contribution to the scattering. This challenge is further amplified within polydisperse systems characterized by a higher number of disks, which may explain the performance deterioration being particularly large for higher numbers of disks. The performance is still generally quite good, with only 15 cases, red circles plus orange circles in Figure 5, out of 1900 showing errors more than one disk compared to true disk numbers, and only one case, orange circle in Figure 5, of 1900 showing errors more than 2 disks.  When delving into the intricacies of polydisperse systems, we noted a decline in precision to approximately 0.6 for configurations involving 19 and 20 disks. While such a precision is arguably acceptable, it prompts a pertinent question: At what point does our machine learning (ML) model's capability to accurately discern disk quantities diminish? Regrettably, due to constraints in both time and resources, we were unable to fully address this dimension of our research. Exploring a broader range of disk quantities, particularly towards more extreme conditions, would be instrumental in assessing the robustness and adaptability of ML models in challenging situations.

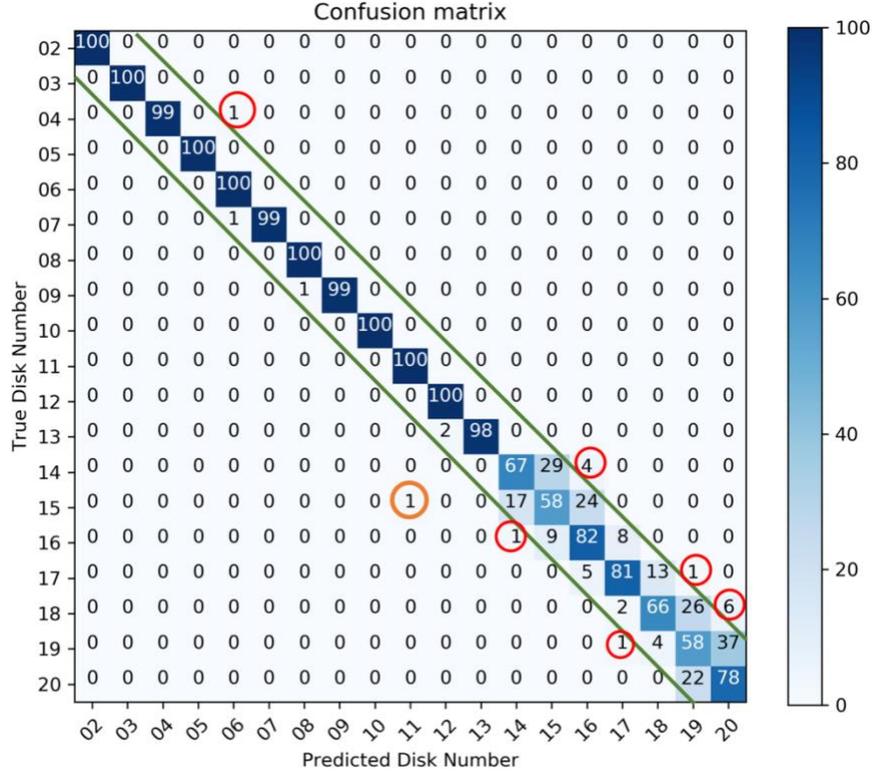

*Figure 5.Confusion Matrix of polydispersity system classification results using $257 \times 257$ as input size of coherent X-ray speckle patterns. The green line illustrates the offset of one disk from the principal diagonal, which symbolizes correct classifications. Consequently, off-diagonal cells within green lines indicate predictions that are merely 1 disk apart from the accurate disk number prediction. The red circle captures the predictions that deviate by 2 disks from the correct disk number prediction, while the orange circle denotes predictions that diverge by more than 2 disks from the correct disk number predictions.*

*Table 1. Classification performance for different polydisperse disk numbers.*

| Disk Numbers | precision | recall | F1-score |
|---|---|---|---|
| Disks_02 | 1 | 1 | 1 |
| **Disks_03** | 1 | 1 | 1 |
| **Disks_04** | 1 | 0.99 | 0.99 |
| **Disks_05** | 1 | 1 | 1 |
| **Disks_06** | 0.98 | 1 | 0.99 |
| **Disks_07** | 1 | 0.99 | 0.99 |
| **Disks_08** | 0.99 | 1 | 1 |
| **Disks_09** | 1 | 0.99 | 0.99 |
| **Disks_10** | 1 | 1 | 1 |
| **Disks_11** | 0.99 | 1 | 1 |
| **Disks_12** | 0.98 | 1 | 0.99 |
| **Disks_13** | 1 | 0.98 | 0.99 |
| **Disks_14** | 0.79 | 0.67 | 0.72 |
| **Disks_15** | 0.6 | 0.58 | 0.59 |
| **Disks_16** | 0.71 | 0.82 | 0.76 |
| **Disks_17** | 0.88 | 0.81 | 0.84 |
| **Disks_18** | 0.8 | 0.66 | 0.72 |
| **Disks_19** | 0.54 | 0.58 | 0.56 |
| **Disks_20** | 0.64 | 0.78 | 0.71 |

## 4. Discussion and Summary

We developed a deep learning based model to identify disk number by classification from coherent X-ray speckle patterns of a two dimensional disk model system. The classifier was tested with and without dispersity and shown to be effective in both cases, although systems with polydispersity show significant errors. Overall, our results demonstrate that without using complex experimental procedures e.g., taking multiple snapshots of samples in different angles or positions for phase retrieval, or pre-defined assumptions about samples, we could directly extract real space sample information such as number density from $k$ space X-ray speckle patterns.

The way we generated the images of $n$-disk systems produced a direct correlation between the average disk size and the average disk number density. Since the average disk size can be estimated from conventional X-ray scattering analysis, we also conducted a comparison of our ML-based analysis with conventional analysis. Conventional X-ray scattering analysis would determine the average disk number density by looking at the overall width of the scattering; smaller disks would give a broader width. We quantify the overall width of the scattering by its first moment, i.e., the average of the product of the pixel intensity and its distance from $q = 0$, typically divided by the average intensity. We determine the first moment on the same speckle patterns used for the ML (specifically, the speckle patterns had central regions removed and intensities standardized, as discussed in Section 2) as well as ones without standardization. Below we will use $513 \times 513$ Non-Polydisperse system, since ML model exhibits good predicting performance for it, to show the relationship between mean first moment and disk number in raw speckle intensity and standardized speckle intensity. Our ML model uses standardization as preprocessing for speckle patterns which makes some intensities zero and some negative and thus creates some uncertainty in how to analyze the first moment of the standardized speckle patterns. Therefore, we will show two different definitions of approximated first moment for standardized speckles. The first definition, which we refer to as using "shifted standardized" data, is just the average of the product of the pixel intensity and its distance from $q = 0$ divided by the average intensity plus one to avoid dividing by zero. The second definition, which we refer to as using "absolute standardized" data, is the average of the product of the absolute value of pixel intensity and its distance from $q = 0$ to make sure negative pixel intensity does not cancel positive intensity. Since in X-ray width analysis, normalized raw intensity to 0 ~ 1 range is often used, we also present normalized intensity results of using Min-Max Scaler, $(I - I_{min})/(I_{max} - I_{min})$, which is referred as "normalized standardized" data. We select 10 speckle patterns for each disk number and show both the raw scatter point plot and the aggregated plot of the mean of those 10 speckle patterns in Figure 6.

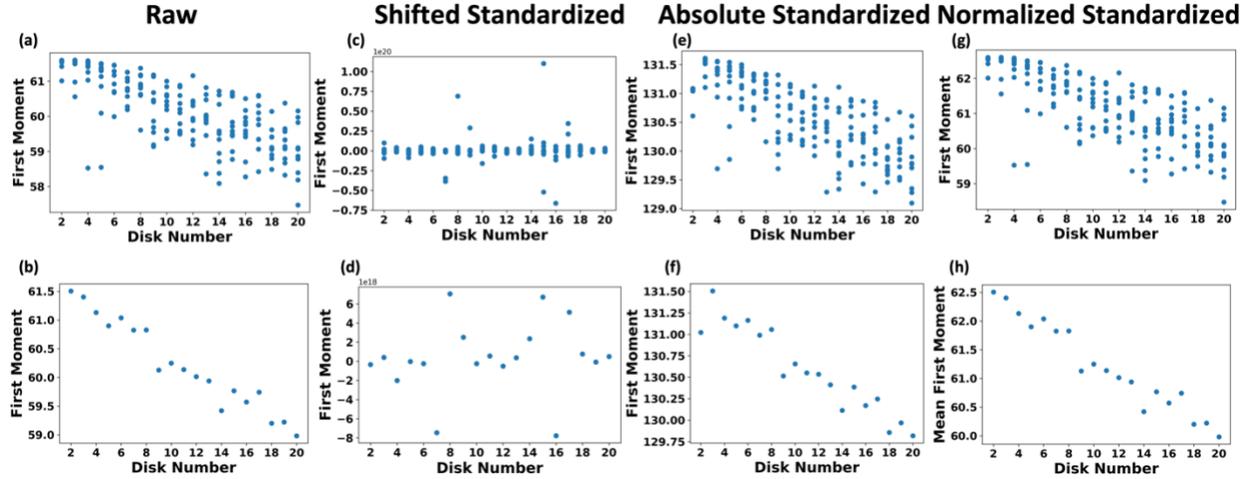

*Figure 6. Relationship between first moment and disk number in raw intensity, shifted standardized, absolute standardized and normalized standardized ML data. (a), (c), (e), (g) show the 10 sampled speckles, and (b), (d), (f), (h) show the mean value of those 10 samples.*

Figure 6(a) shows a clear relationship between mean first moment and disk number. This relationship is largely lost for the shift standardized data (Figure 6(c)) but still quite strong in the absolute standardized data (Figure 6(e)) and normalized standardized data (Figure 6(g)). This result provides evidence that there is indeed relationship between disk number and intensity in the data which the ML can learn. While this analysis shows that first moment and disk number are related, the large scatter in all the cases Figure 6(a,c,e,g), and particularly for shifted standardized data, means that many samples must be taken for a given disk count to robustly estimate the disk count value. However, our ML model can do this with just one speckle patterns, demonstrating it is much more powerful than just the first moment.

It is also of interest to explore those aspects of the speckle patterns are being used by the ML model in determining disk count. To explore this question, we applied Saliency Maps[57] and Gradient-weighted Class Activation Mapping (Grad-CAM)[58], two widely used techniques to analyze CNN based image classification system, which can give insight into the pixel regions that are important for final classifications. Saliency Map serves as a visualization method that identifies the most influential pixels in an input image relative to the output of a CNN. By highlighting these critical pixels, Saliency Map offers a means to discern which aspects of the image most significantly contribute to the model's decision-making process. On the other hand, Grad-CAM provides a more sophisticated visualization by examining the gradients flowing into the final convolutional layer of a CNN. This enables the extraction of high-level spatial information, resulting in a heat map that emphasizes regions of the image vital for the model's predictions. Together, these techniques illuminate the inner workings of CNN-based image classification systems, revealing the specific pixel regions pivotal for their ultimate classifications. Although Grad-CAM and Saliency Map have been used for interpretations of natural images[59,60], photovoltaics[61], small angle X-ray diffraction data[37] and galaxy evolution[62], the results from our ML models overlapping with coherent X-ray speckle patterns did not point us to any clear interpretations of what pixels were being used to identify which categories. We attach results of Saliency Map and Grad-CAM results in SI for further references.

Although the present results are encouraging, significantly more work is needed to assure that machine learning models can be used to aid in interpretation of speckle patterns. First, real coherent X-ray speckle patterns have noise due to environments, devices, detectors etc. The impacts of noise on the classification need to be assessed, although it is likely that this does not represent a fundamental challenge. We expect that enough data with well-enough controlled noise will yield robust models, as found for most of the data here. Second, our model loses accuracy in larger and polydisperse systems, which is likely due at least in part to having many disks with different sizes that blur peaks and thus make it hard to discover patterns. This problem could potentially be reduced by adding more training speckle patterns for larger disk numbers. A closely related third limitation is that it is still unclear what kinds of training are needed to enable extraction of useful data e.g., how many training speckle patterns are needed to train a successful speckle pattern extraction model. Perhaps most importantly, it is not clear how such training data can be practically obtained. The data would likely have to be obtained using systems with ground-truth properties accessible by some other method, perhaps aided by other experiments to find those properties, or simulations where all the properties are readily available. One might consider using data from colloidal systems, which are relatively easy to investigate[63]. We note a fourth limitation which is that our system is a model 2D system. Machine learning approaches should be tested on data of 3D systems and if possible, tested with relevant experimental data. For example, in our investigation, we considered disks with only amplitude (absorption) contrast. However, objects can have not only amplitude contrast but also phase contrast due to refractive index variations. An interesting avenue for future work will be to consider phase objects or those having both amplitude and phase contrast. Despite these many limitations, the success of these initial results suggests that further related studies shall be fruitful.

**Data And Code Availability**

The data described in the Supporting Information and the data used in plotting all figures in manuscript are available at Figshare (https://doi.org/10.6084/m9.figshare.16850398.v4). The Supporting Information data includes Saliency Map Images and Class Activation Map (CAM) Images for both non-polydisperse systems of size $257 \times 257$ and $513 \times 513$, and a polydisperse system with size $257 \times 257$.

The source code for generating the X-ray speckle patterns, training ResNet based classification, and analyzing classification results are available on GitHub ( https://github.com/uw-cmg/ML4XraySpecklePattern ).

**Competing Interests**

There are no competing interests in relation to the work described.

**Acknowledgments**

This material is based upon work supported by Laboratory Directed Research and Development (LDRD) funding from Argonne National Laboratory, provided by the Director, Office of Science, of the U.S. Department of Energy (DOE) under Contract No. DE-AC02-06CH11357. Use of the Center for Nanoscale Materials, an Office of Science user facility, was supported by the DOE, Office of Science, Office of Basic Energy Sciences (BES), under Contract No. DE-AC02-06CH11357. X-ray analysis was supported by DOE BES Materials Sciences and Engineering. We would like to thank the Wisconsin Applied Computing Center (WACC) for providing access to the CPU/GPU cluster, Euler. Special thanks to Colin Vanden Heuvel for helping us use GPUs and install the required software.